\documentclass[usenatbib, twocolumn, numberedappendix, twocolappendix]{openjournal}

\usepackage{newtxtext,newtxmath}
\usepackage[T1]{fontenc}

\usepackage{graphicx}	% Including figure files
\usepackage{amsmath}	% Advanced maths commands
\usepackage[]{hyperref}
\usepackage{xcolor}
\usepackage{soul}

\begin{document}

\title[Dynamical friction and splashback radius]{Dynamical friction and measurements of the splashback radius in galaxy clusters}

% The list of authors, and the short list which is used in the headers.
% If you need two or more lines of authors, add an extra line using \newauthor
\author{Talia M. O'Shea$^{1, 2,3}$}
\author{Josh Borrow$^{4,3}$}
\author{Stephanie O’Neil$^{3, 4, 5}$}
\author{Mark Vogelsberger$^{3}$}
\thanks{$^*$E-mail: tmoshea@wisc.edu}
% List of institutions
\affiliation{$^{1}$Department of Astronomy, University of Wisconsin–Madison, Madison, WI 53706, USA \\
$^{2}$Department of Physics, Wellesley College, Wellesley, MA 02481, USA \\
$^{3}$Department of Physics and Kavli Institute for Astrophysics and Space Research, Massachusetts Institute of Technology, Cambridge, MA 02139, USA\\
$^{4}$Department of Physics and Astronomy, University of Pennsylvania, 209 South 33rd Street, Philadelphia, PA, USA 19104\\
$^{5}$Department of Physics, Princeton University, Princeton, NJ 08544, USA}

\begin{abstract}
The splashback radius is one popular method of constraining the size of galaxy clusters, often measured through the logarithmic derivative of the galaxy number density profile. However, measuring the splashback radius through the galaxy number density has consistently produced smaller values of the splashback radius than those inferred from the underlying gravitational potential in simulations. Dynamical friction has been posited as one possible reason that splashback radii measured through galaxy number densities are reduced, since it decays the orbits of subhaloes within the halo. Dynamical friction is an emergent process, and as such, cannot be isolated or removed within N-body simulations. Here, we present simulations starting with isolated galaxy clusters drawn from the IllustrisTNG cosmological simulation, where we explicitly control dynamical friction through an idealized model. We show that although dynamical friction can reduce measurements of the splashback radius, it does not have a significant effect on clusters with $M_\mathrm{200,mean} > 10^{14} \mathrm{M_\odot}$, and thus cannot account for previously measured discrepancies. \\[0.5em]

\end{abstract}

\maketitle

%%%%%%%%%%%%%%%%%%%%%%%%%%%%%%%%%%%%%%%%%%%%%%%%%%

%%%%%%%%%%%%%%%%% BODY OF PAPER %%%%%%%%%%%%%%%%%%

\section{Introduction}

The environment of the galaxy cluster is a challenging one to understand. Galaxy clusters, the most massive gravitationally bound objects in the Universe, can have masses exceeding 10$^{15}$ M$_{\odot}$ and may contain anywhere from hundreds to thousands of galaxies \citep[henceforth called cluster galaxies; see e.g.][for a review]{Kravtsov2012}. Compared to the rest of the Universe, the cluster environment is dense and hot; the intracluster medium, or ICM, has densities $\sim10^{-3}~\mathrm{cm^{-3}}$ \citep[see e.g.][]{Eckert2012}. The deep gravitational potential well, as well as strong accretion shocks and feedback from active galactic nuclei \citep[AGN; see][]{McNamara2012, Kravtsov2012} heat the gas to temperatures $\sim 10^7 - 10^8$~K, producing X-rays via thermal bremsstrahlung \citep[e.g.][]{Leccardi2008, Chen2024}. Clusters can thus be detected through this X-ray emission \citep[see e.g.][]{Bohringer2001, Pacaud2016}, as well as through the Sunyaev-Zeldovich effect \citep[SZ effect; see e.g.][]{Bleem2015,Planck2016_SZeffect}.

Clusters are also selected and studied through overdensities in their resident galaxies, in the optical \citep[see e.g.][]{Kepner1999, Rykoff2014, Sarron2018, Huang2021} and near-infrared \citep[see e.g.][at $z>1.3$]{Wylezalek2013, Rettura2014,Gully2024}. On average, galaxies residing in clusters do not resemble their counterpart field galaxies (galaxies that do not reside in clusters or groups): they are typically redder and more elliptical \citep{Dressler1980, Hogg2004, Kauffman2004, Blanton2005, Li2006}, and have lower star formation rates \citep{Poggianti1999, Gomez2003, ONeil2023}. Previous studies have sought to understand these differences, notably the rapid cessation of star formation (quenching) in galaxies as they fall into the cluster, identified for instance by the strength of the 4000~{\AA} break \citep{Dressler1987, Kimble1989, Kauffman2004, Li2006, Demarco2010, Bluck2020, Kim2022}. 

One explanation, known as ``environmental quenching'', attributes the lack of star formation in cluster galaxies to a stripping processes that cause galaxies to lose their gas envelopes as they enter the cluster. Ram pressure stripping results from interactions between the hot cluster gas and the gas in an infalling galaxy, which removes gas from the galaxy and thus lowers star formation rates \citep{Gunn1972, Abadi1999, Poggianti1999, Vollmer2004, Chung2009, Ebeling2014}, especially in the center of clusters \citep{Wetzel2013, Borrow2023}. Cluster galaxies are also impacted by tidal stripping, which is a gravitationally driven process in which mass at the edges of an infalling galaxy is pulled off. As a result, infalling galaxies lose mass, including gas. Tidal stripping has been observed in many N-body simulations \citep{Dekel2003, Diemand2007}, and like ram pressure stripping, is thought to operate most efficiently in the center of galaxy clusters \citep{Larson1980, Moore1998}. Subhaloes can also affect each other gravitationally in a process called harassment \citep{Moore1998, Wetzel2013}, or even merge. Additionally, infalling galaxies can be ``pre-processed'' by smaller galaxy groups as they fall into the cluster along filaments \citep{Fujita2004, Sarron2019, Lee2022}.

Yet another process that occurs in the cluster environment is dynamical friction \citep[DF;][]{Chandrasekhar1943,vandenBosch1999}. Dynamical friction is a drag force that acts on objects (e.g. a subhalo) moving through a background comprised of particles (e.g. the cluster dark matter halo). Dynamical friction decays the orbit of the infalling object, with the effect strongest in this context on high-mass galaxies, causing them to merge with the central brightest central galaxy \citep[BCG;][]{Ostriker1977}. \\

A foundational property of a galaxy cluster is its size, and a variety of metrics are used to measure cluster size, such as the virial radius $R_\mathrm{vir}$ assuming a spherical collapse model \citep{Bryan1998}. Based on the spherical collapse model, a related approach for measuring cluster size is to define a radius at which the average enclosed density is some multiple of a reference density \citep[see][]{Cole1996, Diemer2013}. For instance, $R_{200, \mathrm{mean}}$ refers to the radius at which the average enclosed density is two hundred times greater than the mean density of the Universe \citep{Peebles1993, Peacock1999}, which in simulations can be easily calculated for the box volume. The mass enclosed by $R_{200, \mathrm{mean}}$ is accordingly referred to as $M_{200, \mathrm{mean}}$. \citep[Other common analogous radii include $R_{200, \mathrm{crit}}$, $R_{500, \mathrm{mean}}$, and $R_{500, \mathrm{crit}}$, which are more useful for studying the inner regions of haloes, as discussed by][]{Diemer2013}. However, since this metric is defined relative to the Universe as a whole, thus, as the density of the Universe changes over cosmic time, $R_{200, \mathrm{mean}}$ will change along with the Universe to maintain a constant density ratio, a process known as pseudoevolution \citep{Diemer2013}, regardless of the accretion of material. This phenomenon has been measured in numerous N-body simulations where the virial mass increases without a corresponding level of accretion \citep{Prada2006, Diemand2007,Zemp2014, Shirasaki2019, Wang2020}. In particular, \cite{Wetzel2015} found that the dark matter density profile is strongly affected by pseudoevolution. The consequence of pseudoevolution is that the value of $M_{200, \mathrm{mean}}$ changes with time based on both the physical evolution of the cluster, and the evolution of the Universe at large. \\

To avoid pseudoevolution and more accurately trace halo mass accretion histories, other methods of measuring halo boundaries have emerged, for example, the turnaround radius or the depletion radius \citep[see e.g.][respectively]{Pavlidou2014, Fong2021}. In particular, the splashback radius, or $R_{\mathrm{sp}}$, has gained attention in recent years \citep{Adhikari2014}. The splashback radius characterizes the size of the cluster based on the radius of the first apocenter of infalling material, thus probing the mass accretion history \citep{Diemer2014, More2015}. This material produces a ``pile-up'' of matter around that first turn-around point due to the reduced velocity of subhaloes at their apocenters \citep{Adhikari2021}. As the splashback radius is defined with respect to the dynamics of the cluster, it does not experience pseudoevolution but evolves between $\sim$ 1-2 $R_\mathrm{200,mean}$, varying with mass, redshift, and mass accretion rate \citep[see e.g.][]{Diemer2017, ONeil2021, Towler2024}.

The dynamical splashback feature is associated with a caustic in the density profile. In particular, the logarithmic density profile (dlog$\rho$/dlog$r$) exhibits a dip in the profile at this region. The local minimum of that feature marks an excellent proxy for $R_{\mathrm{sp}}$ \citep{Diemer2014}. Due to observational and computational challenges in finding the apocenters of numerous cluster galaxies or particles, the density profile is therefore a commonly used proxy for finding the splashback radius \citep[see e.g.][]{More2015, Chang2018, Shin2019, Murata2020, ONeil2021}. For perfectly spherical systems, this caustic and the apocenter are well-matched \citep{Adhikari2014,Mansfield2017}. However, in reality, this correlation is not perfect. Clusters are not perfectly spherical, which can reduce the strength of the caustic \citep{Diemer2014, Diemer2017, Mansfield2017,Xhakaj2020}. Nevertheless, because of its simplicity, the differential density profile remains the primary way of measuring $R_{\mathrm{sp}}$. In particular, the galaxy number density profile is commonly used because galaxy overdensities are a common tracer of clusters.

Splashback radii have been measured in optically-selected clusters from weak lensing mass measurements \citep[e.g.][]{Chang2018,Fong2022, Giocoli2024, Xu2024} and through galaxy number density \citep[e.g.][]{More2016, Baxter2017,Shin2019, Murata2020,Bianconi2021,Kopylova2022, Contigiani2023}, as well as in X-ray selected clusters, similarly with weak lensing \citep{Umetsu2017,Contigiani2019} and number density of galaxies \citep{Bianconi2021, Rana2023}. SZ-selected clusters have also been studied through galaxy density profiles \citep{Shin2019, Zurcher2019, Adhikari2021, Shin2021}, weak lensing \citep{Shin2021}, and even gas pressure profiles \citep{Anbajagne2022}. \cite{Gonzalez2021} even detected the splashback radius in intracluster light, perhaps also the first detection of the splashback radius for an individual cluster, rather than stacked haloes. 

\begin{figure*}
    \centering
	\includegraphics{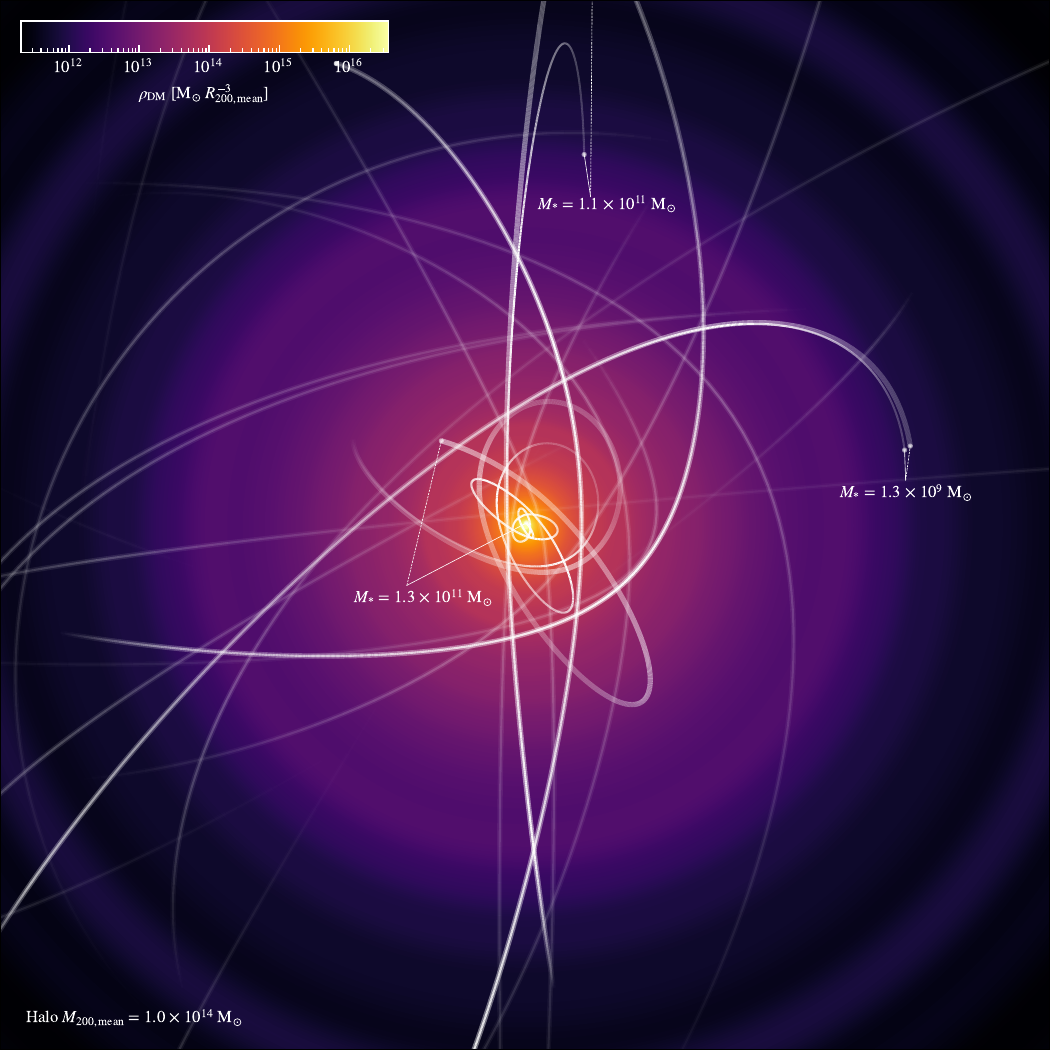}
    \caption{Shows selected galaxy orbits, evolved for 10~Gyr in a cluster. The background colour shows the cluster's spherically averaged dark matter density. Orbits for each galaxy are shown with and without dynamical friction enabled; those with dynamical friction are shown by brighter and thinner lines. The transparency of a line at a given position depends on its distance into the page. Selected galaxy masses are labeled, with solid and dashed lines connecting the labels to the final positions of the galaxies with and without dynamical friction, respectively.}
    \label{fig:prettypicture}
\end{figure*}

In simulation, splashback radii have been detected dynamically through particle orbit tracing \citep[e.g. finding the particle apocenters][]{Diemer2017_rspdetection, Garcia2023, Shin2023}, as well as through dark matter density profiles in N-body and hydrodynamic simulations \citep{Lau2015, Aung2021,Deason2021, ONeil2021, McAlpine2022, Haggar2024, Lebeau2024}. Baryonic contributions have also been studied \citep[e.g.][]{Zhang2023, Towler2024} in gas \citep{Lau2015,ONeil2021, Lebeau2024} and subhalo number densities \citep[see e.g.][all using IllustrisTNG data]{ONeil2021, Dacunha2022, ONeil2022, Pizzardo2024}. 

Cluster splashback radii have been found to depend strongly on the accretion rate of the cluster, which refers to the rate at which the cluster gains mass \citep{Diemer2014, Shin2023}. Haloes with lower accretion rates possess larger splashback radii, relative to $R_{200, \mathrm{mean}}$, than those with higher mass accretion rates \citep{More2015}. \\

Much about the exact relationship between the differential density profile and the splashback radius, as defined from dynamics, remains unknown. Further, even the systematics relating differential density profiles measured from different proxies are poorly understood. Measuring the mass density profile of a halo, which is dominated by dark matter, is challenging in observation; ideally, the galaxy number density profile would act as a useful observational substitute for the dark matter density profile. However, simulations reveal a discrepancy between the dark matter density profile and the galaxy number density profile, affecting the position of the splashback radius. This disagreement has been identified in numerous simulations and observational studies \citep[see e.g.][]{More2016,Bahe2019,ONeil2021, McAlpine2022}.

Some researchers have posited that this discrepancy is due to dynamical friction, largely using simulations \citep[see e.g.][]{Bahe2019, Xhakaj2020, McAlpine2022}. Dynamical friction affects primarily the subhaloes (and the associated galaxies), reducing the extent of their orbits, but not the background dark matter, and consequently reducing the splashback radius measurement from galaxy number density profiles. Using a toy model, \citet{Adhikari2016} simulated dynamical friction in a galaxy cluster system, finding that DF reduced the splashback radius, when measured through galaxy number densities, in a spherical system where all galaxies moved in perfectly radial orbits.

However, some studies argue that dynamical friction's influence is overstated, for example that it is insignificant at splashback scales \citep{Contigiani2019} or affects only a few very massive subhaloes \citep{vandenBosch2016}. Insensitivities to galaxy magnitude in splashback radius measurements \citep[e.g.][]{More2016, Murata2020} also challenge the expected role of DF, with \citet{Dacunha2022} noting that DF cannot explain why increases in galaxy brightness above $i<-19.4$ do not produce additional decreases in the splashback radius. Recently, \citet{ONeil2022} found, using IllustrisTNG, that the discrepancy between splashback radii measured through dark matter and galaxies depends on the absolute subhalo mass, rather than the mass ratio between the subhalo mass and total halo mass (as expected with the effects of DF), ultimately arguing that selection effects among galaxies are more likely the cause of the smaller-than-expected splashback radii.

Results from observational studies are also mixed: \citet{Chang2018} found that high luminosity galaxies traced out smaller $R_{\mathrm{sp}}$, which is consistent with either the expected influence of dynamical friction or galaxy selection effects; however, they also found this effect to be small and within their measurement uncertainty. \citet{Murata2020} did not find consistent trends in splashback radius values with varying magnitude cuts. Measurements of the splashback radius through galaxy number density, with SZ- or X-ray-selected clusters have been consistent with those expected from the dark matter profile \citep[see e.g.,][]{Zurcher2019, Shin2019, Adhikari2021, Shin2021}, indicating that optical selection effects might significantly contribute to the discrepancy \citep[see also discussion in][]{Murata2020}. 

The issue of dynamical friction's influence in galaxy clusters -- particularly on measurements of the splashback radius -- is thus a topic that is often discussed, but has not been clearly isolated. Large volume cosmological simulations naturally solve for \citep[potentially inaccurate, see][]{vandenBosch2018} dynamical friction as a result of N-body dynamics. This means that dynamical friction cannot be separated from the other gravitational forces, and as such there is an extremely limited body of work studying its effects. To target DF, we must run idealised simulations that explicitly model dynamical friction such that it can be removed for comparison purposes.

In this paper, we seek to address this question by evolving simulated galaxy clusters with an analytical implementation of dynamical friction, allowing the role of dynamical friction to be isolated and clearly understood. By using a realistic population of galaxies extracted from the IllustrisTNG simulation suite, we aim to, for the first time, quantify the potential scale of the impact of DF on the splashback radius of galaxy clusters as measured using the galaxy number density profile. Our overarching aim is not to provide specific numerical values for comparisons to observations (as to perform this specific work we must make a number of key assumptions); rather this paper seeks to take the first steps in quantifying whether or not dynamical friction can impact these measurements, and for what range of galaxy and cluster masses. The rest of the paper is structured as follows. Section \ref{sec:methods} outlines our methods, including the implementation of dynamical friction and the simulations (Sec. \ref{sec:methods_sim}). Our results are described in Section \ref{sec:results}, with discussion and conclusions shown in Secs. \ref{sec:discussion} and \ref{sec:conclusions}.

\section{Methods}\label{sec:methods}
In this work we use data from the IllustrisTNG (TNG) cosmological simulations to reconstruct simplified realizations of clusters and subhaloes before evolving the subhalo orbits further forwards in time. We use cluster information from TNG because, although we create an idealized simulation to test DF, the initial phase-space distribution of the subhaloes is crucial for replicating the splashback radius. Creating a realistic phase-space distribution of subhaloes organically is very difficult, and it is easy to inadvertently destroy the splashback radius by removing the evolutionary history, since the splashback radius by definition reflects billions of years of the cluster's evolution \citep[e.g. $R_\mathrm{sp}$ is strongly dependent on accretion rate, see][]{Diemer2014, Shin2023,Towler2024}. These complications are why we use the TNG data instead of creating our own clusters, even if that approach could, in theory, be more flexible. Nevertheless, our methodology for estimating the impact of dynamical friction is relatively insensitive to our initial set-up, as we re-run the same sample twice with dynamical friction ``on'' and ``off'', investigating the differences between the resulting galaxy number density profiles.

The next sections introduce TNG (\ref{sec:methods_tng}) and how (sub)haloes in TNG are identified and selected (\ref{sec:methods_haloes}), before describing the details of our simulation (\ref{sec:methods_sim}) and fitting procedure (\ref{sec:methods_fitting}).

\subsection{IllustrisTNG Simulations}\label{sec:methods_tng}

We utilize the $z=0$ snapshot of the largest-volume IllustrisTNG cosmological simulations, as introduced by \citet{Nelson2018, Pillepich2018} and publicly released in \citet{Nelson2019}. Though the full TNG simulation series includes three box sizes, we use the highest-resolution of the largest volume runs, labeled as TNG300-1. TNG300-1 has a periodic box with side length 302~Mpc, providing a large sample of galaxy clusters. This volume contains $2500^3$ dark matter particles with masses of $5.9\times10^7~\mathrm{M_\odot}$, and $2500^3$ gas particles with target cell masses of $1.1\times10^7~\mathrm{M_\odot}$. The dark matter particles have a gravitational softening length of 1.5~kpc in comoving units for $z>1$ and physical units for $z \leq 1$, whereas the gas cells utilize an adaptive softening length (minimum 0.37~kpc).

TNG is built on the magnetohyrodynamical (MHD) moving-mesh code Arepo \citep{Springel2010}, with an updated version of the Illustris galaxy formation model \citep{Vogelsberger2013}. In the model, gas cells cool through radiative processes and metal line cooling, until they stochastically form stars. As the stars evolve, stellar winds and supernovae return mass, energy, and enriched metals to the gas. Black holes are also included in the simulation; they are seeded in haloes with masses of $1.2 \times 10^6~\rm M_\odot$. TNG updates parts of the Illustris model, including radio mode feedback from AGN \citep[][]{Weinberger2017}, new supernova wind model \citep{Pillepich2018}, and also includes improvements to the numerical convergence of Arepo \citep{Pakmor2016}.

IllustrisTNG follows a cosmology consistent with the \citet{Planck2016} results. All the runs use cosmological parameters of $\Omega_m =\Omega_{dm}+\Omega_b = 0.3089, \Omega_b = 0.0486,~ \rm and ~\Omega_\Lambda = 0.6911$. Additionally, we use $\sigma_8 = 0.8159, n_s = 0.9667$. The Hubble constant is set as $H_0 = 100h$~km~s$^{-1}$~Mpc$^{-1}$, with $h=0.6774$ \citep{Planck2016}. Some runs are dark-matter (gravity) only, and the rest are fully hydrodynamical runs. We utilize the hydrodynamical runs because they contain data about the galaxies around the clusters (e.g. stellar mass).

\subsection{Halo selection and galaxy definition}\label{sec:methods_haloes}
We follow the halo selection criteria of \citet{ONeil2021}, choosing all haloes of $M_\mathrm{200,mean} > 10^{13} \mathrm{M_\odot}$. Beyond the mass requirement, \citet{ONeil2022} additionally remove haloes that are within $10R_\mathrm{200,mean}$ of a more massive halo, preferentially selecting haloes that are isolated and have not been recently disrupted. This filter is important given that we care about the density profiles outside of the cluster center, and because further simulations of the cluster (see Section \ref{sec:methods_sim}) are based on isolated spherical potentials for each halo. From this process we were left with 1401 haloes at $z=0$. However, we further filter the most massive halo because we identify its density profile as having been significantly disrupted by some less massive haloes within $10R_\mathrm{200,mean}$, producing a final sample of 1400 haloes. 

\begin{table}
    \centering
    \renewcommand{\arraystretch}{1.5} 
    \begin{tabular}{ccccc}
    \hline \hline
    log $M_\mathrm{200,mean}/\mathrm{M_\odot}$ &13-13.5 & 13.5-14 & 14-14.5 & > 14.5 \\
    \hline
    $N_\mathrm{halo}$ & 770 & 394 & 185 & 51\\
    $N_\mathrm{subhalo}/10^6$ & 3.29 & 5.65 & 8.52 & 7.88\\
    $N_\mathrm{galaxy}/10^5$ & 3.49 & 6.36 & 9.93 & 9.38
 \\
    \hline \hline
\end{tabular}
\caption{The number of haloes and subhaloes, as well as subhaloes with stellar mass exceeding $10^7 \mathrm{M_\odot}$ (galaxies), from the TNG300-1 snapshot at $z=0$, separated into 0.5-dex bins by halo mass.}
\label{tab:num_halosubhalo}
\end{table}

Table \ref{tab:num_halosubhalo} shows how the 1400 haloes are split by mass, as well as their constituent subhaloes. Subhaloes were identified with the SUBFIND algorithm \citep{Springel2001} and linked with a Friends-of-Friends algorithm using $b=0.2$. Subhalo properties (e.g. position) are defined relative to the center of the halo, which is located at the most bound particle in the halo. Of these subhaloes, we identify a ``galaxy'' as any subhalo with stellar mass greater than $10^7 \mathrm{M_\odot}$.

\subsection{Simulation details} \label{sec:methods_sim}

Using the haloes and subhaloes identified in the TNG300-1 data, we reconstruct the clusters, relying on on version 1.9 of the \textsc{galpy} package as outlined in \citet{Bovy2015}\footnote{\textsc{galpy} is open source and available for use by the community: \url{http://github.com/jobovy/galpy}}. \textsc{galpy}, among other capabilities, provides the ability to build a gravitational potential and integrate orbits within that potential.

For each cluster in the sample, we construct a spherical enclosed mass profile over 100 bins, spaced logarithmically from $0.01$ to 5$R_{\rm 200, mean}$. The profile accounts for the dark matter, gas, and stellar matter in the TNG snapshot. This enclosed mass profile was then interpolated and transformed to a spherically symmetric potential (arbitrary scaling constants were set with \texttt{ro}~=~1~kpc and \texttt{vo}~=~1~kms$^{-1}$). All subhaloes within $15R_{\rm 200, mean}$ were also reconstructed in \textsc{galpy}, treating them as point particles for the gravitational calculation. From the TNG group catalogues, we include the stellar and total bound mass, positions, velocities, and half-mass radii (used for the DF calculation) of subhaloes. Together, this information was used to reconstruct the cluster and integrate orbits using the 5-4 Dormand-Prince integrator (``dopr54\_c'') in \textsc{galpy}. 

The subhaloes are integrated in the static gravitational potential of the cluster for 10~Gyr, with the subhalo positions outputted every 1~Gyr for our convenience (all results in the manuscript are shown at t=5 Gyr or t=10 Gyr). The use of a gravitational potential to model the gravitational forces of the cluster is consistent with \citet{Adhikari2016} and is necessary in order to avoid inherently creating dynamical friction. The simulation was run twice from the same starting conditions. In one case, we run with the gravity from a static cluster potential as the only force, in which case there is no dynamical friction implemented. In the second, we run the \textsc{galpy} integration with both the cluster potential and a dynamical friction force, which is described in the next section. 

Section \ref{sec:disc_limitations} discusses our motivation for this particular methodology in greater detail. In accordance with our goals for this study, we use a variety of simplifying assumption to effectively isolate dynamical friction, at the expense of a physically realistic cluster.

\subsubsection{Analytical implementation of dynamical friction}\label{sec:methods_df_galpy}

Dynamical friction is implemented in \textsc{galpy} using an analytical approximation based on Chandrasekhar's dynamical friction formula \citep{Chandrasekhar1943} and closely following the implementation of \citet{Petts2016}. The specific representation that \textsc{galpy} uses is written as:
\begin{align}
    \boldsymbol{a}(\boldsymbol{x}, \boldsymbol{v}) = -&2\pi G^2 M \rho(\boldsymbol{x}) \ln \left(1 + \Lambda^2 \right) \cdot
    \\ \nonumber
    & \left( \mathrm{erf}(\boldsymbol{x})  - \frac{2X}{\sqrt{\pi}} \exp \left( - X^2 \right) \right) \frac{\boldsymbol{v}}{v^3}.
    \label{eq:cdf_def}
\end{align}
This expression gives the acceleration of an object of interest, with mass $M$, position $\boldsymbol{x}$, and velocity $\boldsymbol{v}$, moving through a background medium of density $\rho$. The term $X$ is written by,
\begin{equation}
    X = \frac{v}{\sqrt{2} \sigma_r(r)},
    \label{eq:X_def}
\end{equation}
where $\sigma_r$ is the radial velocity dispersion.

One term of Chandrasekhar's formula whose execution frequently varies is the Coulomb logarithm term $\rm ln \left( \Lambda \right)$. $\Lambda$ is related to the ratio of impact parameters $b_{\rm min} / b_{\rm max}$, but writing this quantity in terms of system quantities is difficult (see \citet{Just2005} for a discussion of the complexity). In Appendix \ref{sec:appendixb} we investigate the role of the $\Lambda$ term, but otherwise use the \textsc{galpy} default of:
\begin{equation}
    \Lambda = \frac{r/\gamma}{\mathrm{max} \left(r_{\mathrm{hm}}, GM/v^2 \right)},
    \label{eq:galpy_lambda}
\end{equation}
where $\gamma$ is a scaling term, $r$ is the radial distance of the subhalo from the center at any given point, and $r_{\mathrm{hm}}$ is the half-mass radus of the subhalo. We use the default $\gamma = 1$. For the vast majority of galaxies in our simulation, $r_{\rm hm} > GM/v^2$. 

For the simulations run with dynamical friction, the acceleration is recalculated at every numerical timestep. Each subhalo's half-mass radius and total mass is used in calculating the acceleration, but these are held constant and not updated throughout the simulation.

From Equation (1), we expect dynamical friction to reduce the measured splashback radius for high-mass galaxies, particularly in lower-mass clusters, as well as in denser environments, or those with larger $\Lambda$. We investigate primarily the role of cluster and galaxy mass through our main analysis, with a discussion of the Coulomb logarithm in Appendix \ref{sec:appendixb}.

\begin{figure}
    \centering
    \includegraphics{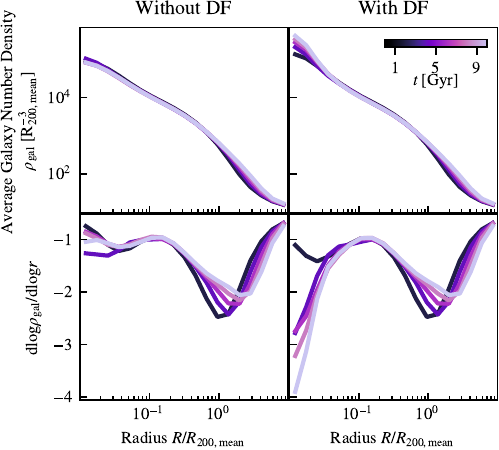}
    \caption{Time evolution of average galaxy number density profiles (upper) and differential density profiles (lower) through 10~Gyr of evolution, as shown in the colourbar. The profiles are evolved with DF disabled (on left) and enabled (on right). Haloes are stacked with $M_{200, \mathrm{mean}}$ between $10^{14}$ and $10^{14.5}~\mathrm{M}_{\odot}$ (185 haloes), and include all subhaloes.}
    \label{fig:profile_and_diff_dens_14.145}
\end{figure}

\subsection{Bootstrapping and fitting}\label{sec:methods_fitting}

We create galaxy number density profiles for each cluster (see Appendix \ref{sec:appendixa} for more discussion). This entails binning the galaxies into 32 bins spaced logarithmically from $10^{-1} - 10^{0.8}R_\mathrm{200,mean}$. The count of galaxies in each bin divided by the volume of that bin produces the density profile.

We then take the logarithmic derivative of these profiles, where the minimum of that profile is the splashback radius. To measure the splashback radius in a robust way and quantify uncertainty, we bootstrap each set of clusters and (sub)set of subhaloes 1000 times, with 32 sub-bootstraps. This process returns bootstrapped density profiles and logarithmic density profiles, as well as errors.

To ensure consistent normalization as we move between cluster masses, we re-scale all of the density profiles before bootstrapping. As reference, we use our sample with halo mass between $10^{14}$ and $10^{14.5}$~M$_\odot$ and with minimum subhalo stellar mass cut of $10^{7}$~M$_\odot$, and no dynamical friction, since this sample hosts the largest number of galaxy-mass subhaloes (see Table \ref{tab:num_halosubhalo}). We compute the average density profile for the reference sample and the sample of interest, and find an average scale factor between them. We then multiply all profiles by that factor, while preserving the shape of the density profile. Doing so reduces spurious trends caused by normalization changes.

The 1000 bootstrapped profiles are fit with open-source fitting code from \citet{ONeil2022}, using fits based on theoretical density profiles from \citet{Diemer2014}, where the overall density is comprised of an inner, outer, and transition region:
\begin{equation}
    \rho(r) = \rho_{\rm inner} \times f_{\rm trans} + \rho_{\rm outer}.
    \label{eq:diemer_overalldensprof}
\end{equation}
The above components can be written as:
\begin{equation}
    \rho_{\rm inner} = \rho_{\rm s} \exp\left(-\frac{2}{\alpha} \left[\left(\frac{r}{r_{\rm s}}\right)^\alpha - 1 \right] \right), 
    \label{eq:diemer_innerdensprof}
\end{equation}
\begin{equation}
    f_{\rm trans} = \left[ 1 + \left(\frac{r}{r_{\rm t}}\right)^\beta \right]^{-\frac{\gamma}{\beta}} 
    \label{eq:diemer_transdensprof}
\end{equation}
\begin{equation}
    \rho_{\rm outer} = \rho_{\rm m} \left[b_{\rm e}\left(\frac{r}{5R_{200,\rm mean}}\right)^{-s_e} + 1 \right], 
    \label{eq:diemer_outerdensprof}
\end{equation}
We adjust the fitting boundaries from Appendix B of \citet{ONeil2021} so that we fit $\rho_{\rm inner}$ at radii  $0.2 < r/R_{200, \rm mean} < 1.2$, with outer fit extending from $2.0 < r/R_{200, \rm mean} < 10.0$. The transition region is fit from $1.2 < r/R_{200, \rm mean} < 2.4$, so there is some overlap between the transition and outer regions during fitting. The lack of mergers in our simulation thus does not affect the splashback radius measurement as the minimum inner radius for fitting is $0.2R_\mathrm{200,mean}$. See Appendix \ref{sec:appendixa_fitting} for further discussion of these regions. 

\section{Results}\label{sec:results}

Figure \ref{fig:prettypicture} illustrates the wide variation in how the orbits of galaxies are affected by dynamical friction. Certain galaxies are greatly affected, spiraling into the dense cluster center, while others experience only moderate changes to their orbits. Many seemingly do not undergo any notable changes, such as the galaxies shown more faintly in the background. Galaxies with properties leading to higher drag forces (galaxies that approach the cluster center closely, are more massive, have higher values of $\Lambda$, etc.) are shown to be more affected by DF. For example, in Figure \ref{fig:prettypicture}, given the labeled stellar masses, the highlighted galaxies have masses approaching or exceeding the widely used $M_{\mathrm{subhalo}}/M_{\mathrm{halo}}\gtrsim 1\%$ metric, where DF is expected to become influential \citep{Adhikari2016, More2016}.

These examples illustrate the qualitative impact of dynamical friction on individual galaxy orbits, but to understand how dynamical friction affects our inference of galaxy cluster properties, like the splashback radius, we must expand our analysis to a large sample. The splashback radius is marked by the minimum in the dip of the differential density profile, as seen in the bottom panels of Figure \ref{fig:profile_and_diff_dens_14.145}, which shows the time evolution of galaxy number density and logarithmic density profiles with and without dynamical friction, summed over all 185 clusters in the mass bin $10^{14} < M_\mathrm{200,mean}/\mathrm{M_\odot} < 10^{15}$ (see Table \ref{tab:num_halosubhalo}). Measurements of $R_{\rm sp}$ are defined relative to $R_{200, \mathrm{mean}}$, because the outskirts of clusters are most self-similar when stacked by $R_{200, \mathrm{mean}}$ \citep{Diemer2014, Lau2015,Shi2016, Umetsu2017}. 

The density profile does not evolve dramatically over 10~Gyr, but the differential density profile does exhibit more significant changes with time evolution. The gradual outward creep of the splashback radius, regardless of DF implementation--growing by approximately a factor of two throughout the simulation--does not represent a physical change in the cluster. Because we do not model continuous accretion of mass in our potential, the value of $R_{200, \mathrm{mean}}$ used here is fixed at its initial value from $z=0$. As shown in \citet{ONeil2023} and \citet{ONeil2024}, the dip in the differential density profile is created by galaxies with infall times typically over 5~Gyr ago, leading to some small offsets in our simulated properties. As we are generally only concerned with the influence of dynamical friction, we are still able to compare the two simulations to each other and so this shift is not problematic. 

With DF enabled, there is a visible overdensity at the center ($\leq0.1R_{200, \mathrm{mean}}$), that grows with time, which is consistent with the orbital decay that dynamical friction produces. In a real cluster, these galaxies may merge with the central cluster galaxy \citep[e.g.][]{Ostriker1977}; however, this simulation does not have a mechanism for mergers, and therefore these galaxies continue to orbit close to the center. Still, the influence of dynamical friction on the splashback region ($\sim R_{\mathrm{200,mean}}$) is not immediately apparent.

\begin{figure}
    \centering
    \includegraphics{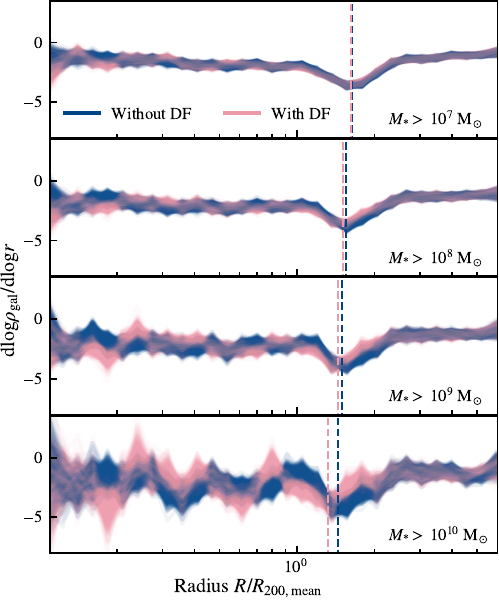}
    \caption{Bootstrapped, unfitted differential density profiles for haloes of $M_{200, \mathrm{mean}}$ between $10^{13}$ and $10^{13.5}~\mathrm{M}_{\odot}$ evolved for 5~Gyr, among subhaloes with minimum stellar mass cuts, as labeled in each panel. Profiles evolved without DF are shown in blue, and those with DF are shown in pink. Vertical dashed lines mark the respective median splashback radius, in matching colours.}
    \label{fig:unfitted_densprofiles}
\end{figure}

\begin{figure}
    \centering\includegraphics{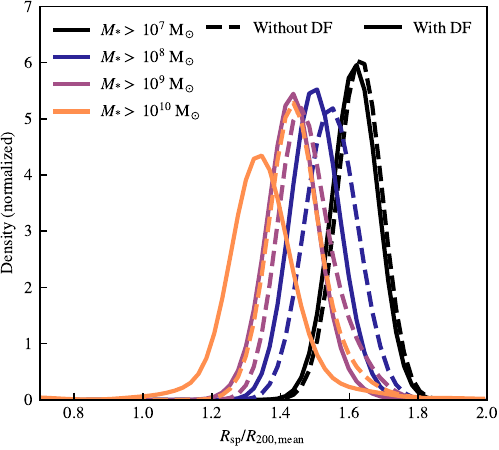}
    \caption{Splashback radius distributions after 5~Gyr for haloes with masses of $10^{13}~<~M_{200, \mathrm{mean}}/\mathrm{M}_{\odot}~<~10^{13.5}$. Minimum stellar mass cuts are shown in different colors. Distributions without dynamical friction are marked with dashed lines, while those evolved with dynamical friction use solid lines.}
    \label{fig:posterior_rsp_distribution_13135}
\end{figure}

\begin{figure}
    \centering
    \includegraphics{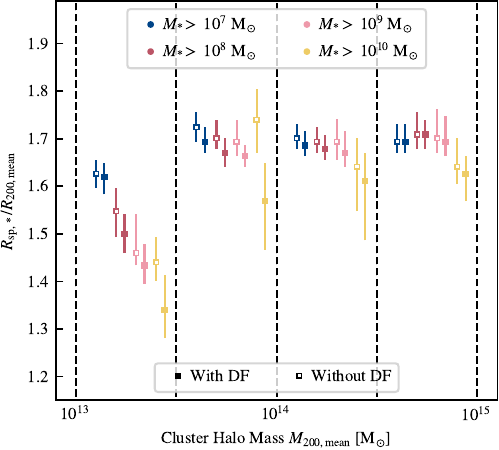}
    \caption{Splashback radius distributions after 5~Gyr from bootstrapped and fitted profiles, divided by halo mass groups of 0.5-dex (separated by vertical black dashed lines.) Within each halo mass division, horizontal position is unconnected to mass. Minimum subhalo stellar mass cuts are individually coloured. Median splashback radii calculated from haloes evolved without DF are shown with unfilled square markers, while those evolved with DF are shown with filled squares. The error bars show the 16-84 percentile of the range of each distribution.}
    \label{fig:rsp_distributions}
\end{figure}

\begin{figure}
    \centering
    \includegraphics{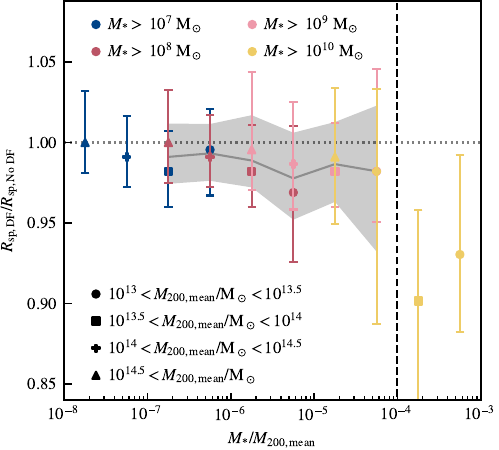}
    \caption{The ratio of the splashback radius with and without dynamical friction, after 5~Gyr, for the ratio of minimum stellar mass to halo mass (measured in the middle of the bin). The horizontal grey dotted line marks a ratio of 1 (i.e. no difference). The vertical black dashed line shows where $M_{\mathrm{subhalo}}/M_{\mathrm{200, mean}} \approx 0.01$. Finally, the solid grey line and shaded band show average ratio and error for the regions with multiple data points.}
    \label{fig:rsp_ratio_plot}
\end{figure}

To quantify the splashback radius, we examine the bootstrapped density profiles for stacked clusters, as described in Section \ref{sec:methods_fitting}. We aim to study the cluster after one crossing time has passed, which we require for the splashback radius to `update' without dynamical friction. The cluster crossing time can be approximated as the mass-independent quantity $t_\mathrm{cross} = t_\mathrm{H}/(5\sqrt{\Omega_\mathrm{m}})$ \citep{Diemer2017_rspdetection}, which evaluates to 5.2~Gyr, following the cosmology of \citet{ONeil2024} Section 3.4. Figs. \ref{fig:unfitted_densprofiles} through \ref{fig:rsp_ratio_plot} thus show our analysis after 5~Gyr of evolution.

Throughout, we divide the clusters into 0.5-dex mass bins for $\log M_{200, \mathrm{mean}}/\mathrm{M_\odot}$ in the ranges 13.0 - 13.5, 13.5 - 14.0, 14.0 - 14.5, and above 14.5. We also divide subhalo galaxies by 1~dex minimum stellar mass bins.

The strength of dynamical friction, as a result of many gravitational interactions, is determined by the subhalo's total mass rather than its stellar mass. However, throughout our work we present results as a function of stellar mass rather than total mass. We do so because in observation, the limiting factor about whether a galaxy is detected is more closely related to its stellar mass than its total mass. Therefore, dividing our galaxy sample by stellar mass allows our simulation to be more easily connected to observational constraints. Additionally, stellar mass is not affected by tidal stripping to the same degree as halo mass \citep{Smith2016}, and we do not have a reasonable stellar evolution model for these galaxies (as we do not track star formation, etc.), making the stellar mass the best proxy for observations. In particular, we follow the stellar mass minimum cuts used in \cite{ONeil2022} in order to be consistent with their analysis, which was based on IllustrisTNG data. Like \citet{ONeil2022}, we include only minimum mass cuts, rather than both maximum and minimum masses, as an analogue to survey magnitude limits.

Although the dynamical underpinnings of the splashback radius refer to the first particle apocenter, the use of the logarithmic density profile by definition always includes particles which are not tracing out their first apocenter. Nevertheless, the position of the splashback radii from the two methods of measurements correspond \citep[setting aside the asphericality of clusters, which is addressed via stacking; for further discussion see e.g.][]{Diemer2014, Adhikari2014, Mansfield2017}.

Figure \ref{fig:unfitted_densprofiles} shows the bootstrapped differential density profiles for the lowest cluster mass bin, with increasing stellar mass cuts for the subhaloes. The profiles are shown before fitting, producing increased noise -- with increased scatter visible as the subhalo sample size shrinks. Even so, Figure \ref{fig:unfitted_densprofiles} visually demonstrates the impact that dynamical friction has on the differential density profiles. Where the profiles appear more blended, dynamical friction has had a weaker effect on the profile. The dip corresponding to the splashback radius is clearly visible, and matches well with the vertical dashed line that identifies the median splashback radius for the sample (post-fitting), such that our visual interpretations here relate well to the quantitative discussion later on. 

\begin{table}
    \centering
    \renewcommand{\arraystretch}{1.5}
    \begin{tabular}{ccccc}
    \hline \hline
    $M_*/\mathrm{M_\odot}$ & $10^7$ & $10^8$ & $10^9$ & $10^{10}$ \\
    \hline
    $\bar{M}_\mathrm{subhalo}/\mathrm{M_\odot}$& 1.5 $\times 10^{10}$ &5.6 $\times 10^{10}$  & 1.6 $\times 10^{11}$ & 4.8 $\times 10^{11}$ \\
    \hline  \hline
    \\
    \hline \hline
    $M_\mathrm{subhalo}/\mathrm{M_\odot}$ & $10^{10}$ & $10^{11}$ & $10^{12}$ & $10^{13}$ \\
    \hline
    $\bar{M}_*/\mathrm{M_\odot}$& 0 &1.7 $\times 10^{8}$  & 2.1 $\times 10^{10}$ & 1.2 $\times 10^{11}$ \\
    \hline \hline

\end{tabular}
\caption{The top two rows show median (barred quantity) subhalo total mass among the entire selected sample of TNG300-1 haloes for the listed stellar masses. The bottom rows show the inverse, with median stellar mass among our cluster sample for given subhalo masses.}
\label{tab:stellar_halomass}
\end{table}

At less massive stellar mass cuts ($M_*>10^7~\mathrm{M}_\odot$) the profiles with and without DF are highly blended. Greater subhalo minimum stellar mass cuts produce increased divergence between the simulations run with and without dynamical friction. As the stellar mass cut increases, both minima move inward (which will be discussed further later), but the profiles with dynamical friction are more significantly shifted towards smaller radii. Notably, this plot examines only the lowest-mass clusters, for which we see the greatest influence of dynamical friction.

After fitting, we extract values of the splashback radius from the minima of each fitted logarithmic density profile, producing posterior distributions such as the one shown in Figure \ref{fig:posterior_rsp_distribution_13135}, which shows a histogram of splashback radius distributions for the lowest mass clusters. That the recorded splashback radii trace a normal distribution indicates that our fitting procedure is stable (e.g. there are no significant outliers and the distribution is not bivariate). As expected, dynamical friction reduces the splashback radius, significantly among high mass galaxies, though barely when all galaxies are included (the black curves).

We are interested in how these trends persist across cluster masses. Figure \ref{fig:rsp_distributions} shows splashback radius distributions, separated by cluster halo mass and subhalo stellar mass. DF reduces the splashback radius prominently among subhaloes of higher stellar mass and for clusters of lower mass (<$10^{14}$ M$_\odot$). These results are consistent with the qualitative trends expected of dynamical friction, with high-mass subhaloes in low-mass haloes primarily affected.

Figure \ref{fig:rsp_distributions} again shows variation in splashback radius as a function of increased galaxy stellar mass, even without dynamical friction (as seen most clearly in the dashed lines of Figure \ref{fig:posterior_rsp_distribution_13135}). This trend, which broadly matches those observed in IllustrisTNG data by \citet{ONeil2022}, still exceeds 1$\sigma$ between $M_*>10^7$ and $M_*>10^{10}$ for the lowest-mass haloes in our sample ($10^{13}<M_{\mathrm{200,mean}}/\mathrm{M_\odot} < 10^{13.5}$). Since our simulations are idealized and, other than dynamical friction, only include gravitational forces between the cluster potential and individual subhalo, we attribute this residual trend to fundamental differences in the phase-space characteristics of massive galaxies, which are stronger in the less massive clusters. Section \ref{sec:disc_residuals} discusses this trend in greater detail.

Though these variations in $R_\mathrm{sp}$ are interesting, in this study, we are primarily interested in the discrepancies between the simulations run with and without dynamical friction. To this end, Figure \ref{fig:rsp_ratio_plot} shows the ratio of the splashback radii with and without dynamical friction, but as a function of minimum subhalo stellar mass over halo mass. By looking at the overlapping points, we can see that despite the residual trends seen in Figs. \ref{fig:posterior_rsp_distribution_13135} and \ref{fig:rsp_distributions}, there is very good agreement when we examine the fraction $R_\mathrm{sp, DF}/R_\mathrm{sp, No DF}$. Therefore we conclude that any contamination of our sample does not significantly affect our analysis of dynamical friction.

It is only after this mass fraction ($M_*/M_\mathrm{200,mean}$) exceeds $10^{-4}$ that we see notable decreases in the splashback radius ($\lesssim10\%$) due to DF, which is consistent with our estimates and the figure produced by \citet{Adhikari2016}. We also note that, as expected, the mass fraction is what drives the trend here, not only the subhalo mass cut or cluster mass. However, \citet{ONeil2021} showed that in the TNG simulations it was the absolute subhalo mass that matters for the decrease in splashback radius, again indicating that dynamical friction cannot be the only reason for those offsets. This difference is particularly of note since we use the same sample as \citet{ONeil2021}.

\subsection{Residence Time}\label{sec:results_restime}

The splashback radius is traced out by subhaloes that have spent sufficient time in the cluster to reach their first apocenter. In position-radial velocity phase-space, then, $R_\mathrm{sp}$ should appear as an overdensity of subhaloes with low radial velocities. Moreover, galaxies that are just infalling into the cluster should not trace out the splashback radius. In this section, we investigate how the cluster phase-space evolves with residence time in our idealized simulations.

\begin{figure*}
    \centering
	\includegraphics{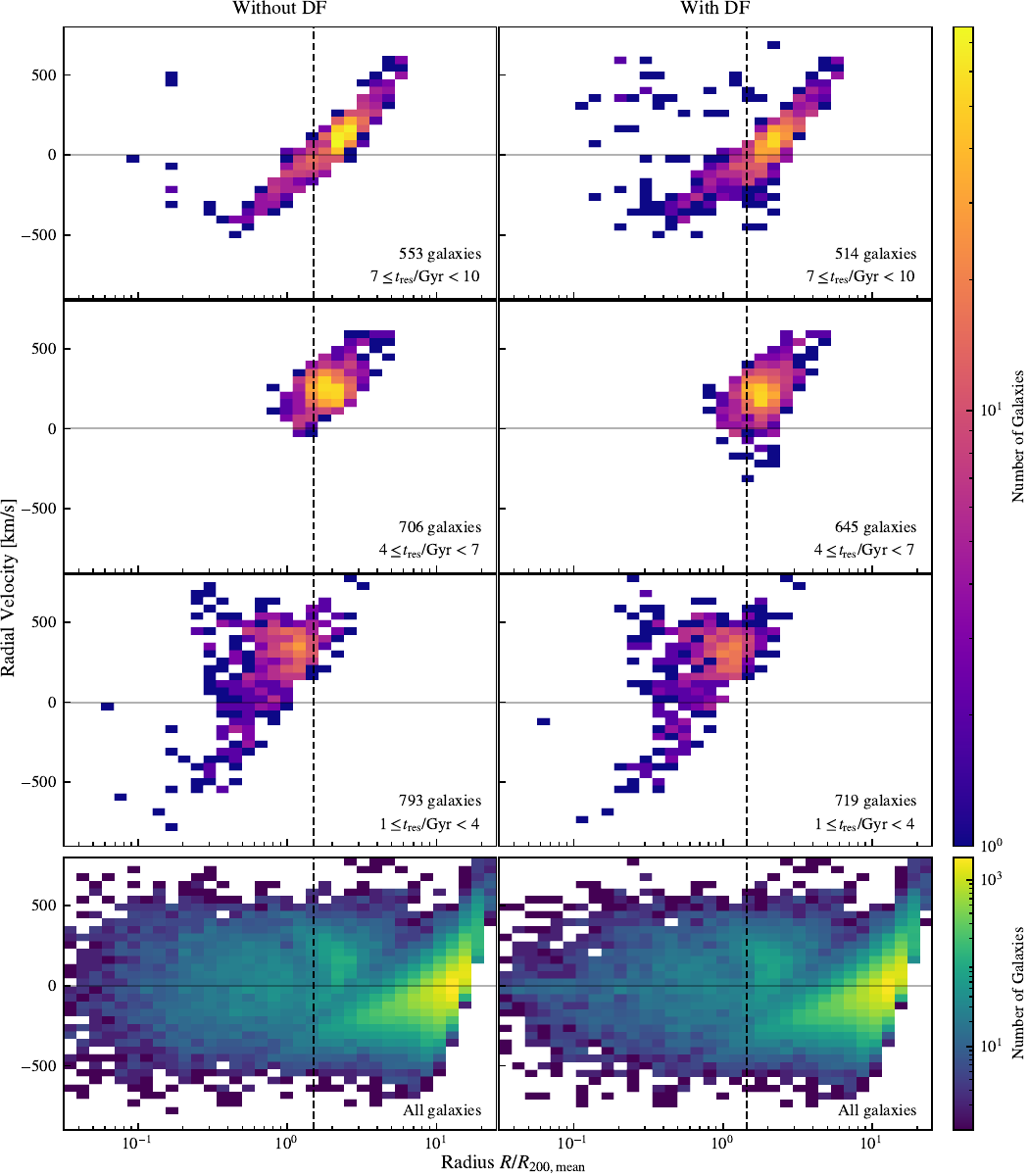}
    \caption{2D histograms of radial velocity and radius plots after 10~Gyr of evolution, with (right column) and without (left) dynamical friction, for the lowest mass haloes ($10^{13}$ - $10^{13.5}$~M$_{\odot}$, totaling 770 haloes) and subhaloes with minimum stellar mass $M_*~=~10^9$~M$_{\odot}$. Subhaloes with $M_{\rm subhalo}~>~0.05 M_{200, \mathrm{mean}}$ are also excluded. The top three panels are separated by residence time in the cluster, which is calculated from the first time at which the galaxy reaches a position within $R_{200, \mathrm{mean}}$. The bottom row shows the histogram for all subhaloes with $M_*$ > 10$^9$~M$_{\odot}$ and $M_\mathrm{subhalo}~<~0.05~M_{200, \mathrm{mean}}$, totaling over 47,000 galaxies. The black dashed line shows the median splashback radius with and without dynamical friction (as measured in Figure \ref{fig:rsp_distributions}.)}
    \label{fig:radialvelocities}
\end{figure*}

The phase-space distributions for high-stellar mass galaxies ($M_* > 10^9~\mathrm{M_\odot}$) are shown in Figure \ref{fig:radialvelocities} after 10~Gyr of evolution, for our lowest-mass clusters (between $10^{13}$ and $10^{13.5} \mathrm{M_\odot}$). We examine the phase space after 10~Gyr, rather than 5~Gyr like in the rest of the analysis, to see longer-duration changes in the cluster environment. We separate the subhaloes by residence time, which is defined as the time since the subhalo first crossed inside of $R_\mathrm{200,mean}$. The top three panels show subhaloes with residence times of 1 to 3~Gyr, 4 to 6~Gyr, and 7 to 10~Gyr. The bottom panel shows all galaxies meeting the mass cut, including galaxies that never reach within $R_{200, \mathrm{mean}}$ or galaxies that began the simulation inside $R_{200, \mathrm{mean}}$. We filter out a small number of galaxies with $M_\mathrm{subhalo} > 0.05 M_{200, \mathrm{mean}}$, as they should, in a physically realistic environment, influence the movement of the cluster. On the plot they would appear as a horizontal band throughout a range of radial positions with very low radial velocities (approximately zero). 

If instead a minimum stellar mass cut of $10^7~\mathrm{M}_\odot$ is applied, the phase-space profiles resemble the expected distribution from IllustrisTNG outputs \citep[see e.g.][]{Dacunha2022, ONeil2024}, suggesting our model reasonably recovers the phase-space measured by other simulated clusters (e.g. the large number of infalling galaxies with negative radial velocity, the dispersal of galaxies at long residence times), even after 10~Gyr of idealized simulation. For galaxies with residence times between 1 and 3~Gyr, we see the infalling and outgoing galaxies combined, as well as a turnaround in phase space around $\sim 5$~Gyr. Thus, we conclude that our simplified clusters (e.g. treating subhaloes as point masses) maintain the expected phase space from N-body simulations. 

The splashback feature is thought to be found among galaxies with residence times of order one crossing time \citep[$\sim$ 5~Gyr, from][]{ONeil2024}, with radial velocities around 0. We begin to see these tracer galaxies in the middle residence time bin, and the number grows with residence times of over 7~Gyr (the top panel). While one might expect the splashback feature to be traced by galaxies with residence times of closer to 5~Gyr, the feature is in fact also visible among galaxies that have spent longer in the cluster (residence time between 7 and 9~Gyr). We attribute this phenomenon to weaker gravitational forces in the least-massive clusters producing less plunging orbits. 

The majority of galaxies in this halo sample never reach inside of $R_{200, \mathrm{mean}}$ (approximately 88\% whether with or without DF.) Of the galaxies that do, the influence of dynamical friction is most apparent among the galaxies with longer residence times. There are few differences between the phase-space distributions for galaxies with $t_\mathrm{res} < 4$~Gyr. By the longest residence time bin, though, the splashback radius (marked with a vertical dashed line) is slightly reduced. The distribution of radii at $v_r \approx 0$ is wider with dynamical friction enabled as well, matching the wider bootstrapped $R_{\mathrm{sp}}$ distributions shown in Figure \ref{fig:violinplot}. 

A small number of galaxies with $v_r < 0$ and $4 \leq t_\mathrm{res} < 7$ appear only with dynamical friction enabled. By comparison to the phase-space distribution with a lower mass cut, it is apparent that (essentially) only galaxies with high masses exist in this region of phase space, and only when dynamical friction is acting. We can thus see how DF, for a small number of high mass galaxies, dramatically affects their orbital properties (e.g. as visualized in Fig. \ref{fig:prettypicture}). However, despite these cases, it is also clear that the phase-space distributions are not, overall, dramatically altered with dynamical friction enabled. 

\section{Discussion}\label{sec:discussion}
This work is motivated by a lack of studies specifically targeting the role of dynamical friction in the galaxy cluster context. As far as the authors are aware, only \citet{Adhikari2016} specifically isolates DF in their toy model simulations, finding that DF reduced the splashback radius, when measured through galaxy number densities, in a spherical system where all galaxies moved in perfectly radial orbits. \citet{Adhikari2016} concluded not only that dynamical friction resulted in a noticeable decrease in $R_{\mathrm{sp}}$ for subhaloes above 1\% of the host halo mass, but also that the influence of DF is dependent on the cluster potential profile. We are motivated to create a complementary study. 

\subsection{Limiting assumptions}\label{sec:disc_limitations}

Practically speaking, isolating DF requires numerous simplifying assumptions about the cluster environment, which forms and evolves through the intersection of many processes (outlined briefly in the Introduction). The initial conditions of our simulation (i.e. the $z=0$ TNG snapshot) evolved with dynamical friction, thus impacting the orbits of the galaxies we evolve further. Although we allow one crossing time to pass before analyzing the location of $R_{\rm sp}$, we cannot fully quantify the impact of our initial conditions on our findings, whether that might produce an overestimate (e.g., due to pre-existing mass-segregation) or underestimate in the role of DF. However, as both the with- and without- DF simulations start from identical snapshots, we believe it is still useful to compare between the runs. We stress that our results, therefore, \textit{do not aim to accurately calculate values of the splashback radius, but instead to gain a preliminary upper limit on how $R_\mathrm{sp}$ measurements might be affected by the presence of dynamical friction}. This includes our use of a static spherical gravitational potential for the clusters and the lack of baryonic physics applied to the subhaloes -- each of which we discuss in the Methods section. More complexity (e.g. the inclusion of baryonic physics) could be added in future studies, as this work is only a first step towards more rigorously understanding processes that impact measurements of the splashback radius and other proxies that rely on the galaxy number density profile.

The use of a gravitational potential rather than N-body calculations is necessary in order to avoid creating dynamical friction. We also intentionally exclude accretion in order to isolate dynamical friction. The unmodeled growth of the halo potential with time may cause an underestimate of DF. However, work by \cite{Diemer2013, Wu2013} indicates that most of the cluster mass growth from $z=1$ ($>5$~Gyr ago) until the present day is due to pseudoevolution, rather than physical accretion of mass. This minimal real mass accretion should be particularly negligible and continue into the simulated future for our sample of relaxed, isolated clusters.

We treat our subhaloes as point masses with fixed mass traversing a spherical gravitational potential. Both of these choices are not strictly accurate, as clusters are elliptical and galaxies lose mass through stripping as they enter the cluster. Nevertheless, both of these choices are consistent with the methodology used in \citet{Adhikari2016}. The treatment of clusters as spherically symmetric follows that of \cite{Adhikari2014}, who find that the use of spherical cluster potentials does not substantially alter their measurement of the splashback radius \citep[see also][for further discussion of sphericality of clusters]{Olamaie2012}. Furthermore, whether observed or simulated, all downstream calculations of the differential density profile in the literature assume a spherical halo profile, consistent with our methodology. In fact, as the subhalo velocities and properties are drawn from TNG data, the subhalo phase space, and the direction of infall, is improved relative to the radial-only movement of the subhaloes in \citep{Adhikari2016}. 

Our exclusion of mass stripping is necessary to reduce confounding factors in our study of dynamical friction alone. As both ram-pressure and tidal stripping are dependent on the distance from the cluster center, subhaloes may experience different mass loss rates if their orbits are decayed by DF. In addition to potentially confounding our results, incorporation of baryons into our simulations would significantly expand the scope and computational requirements of this work, without commensurate improvements to our study of dynamical friction in isolation. Even so, we attempt to ameliorate errors from these choices by selecting galaxies by stellar mass throughout our analysis, which is significantly more stable against stripping than subhalo mass \citep[see e.g.][]{Penarrubia2008, Smith2013, Smith2016}. More importantly, any decrease in the subhalo mass will reduce the impact of dynamical friction, meaning that our results can be seen as an upper limit on how significantly DF impacts the splashback radius.

Our comparisons to other works (e.g. observations and cosmological simulations) of the splashback radius should be read with these limitations in mind. That our phase-space resembles that of IllustrisTNG (see Section \ref{sec:results_restime}) is a reassuring sign that we can compare our results to these other studies. Splashback radius studies typically stack clusters over redshift bins corresponding to spacings of $\gtrsim 2-3$~Gyr \citep[see e.g. those of][]{More2016,Chang2018, Murata2020}, and furthermore, find that the splashback radius scales with $R_\mathrm{200,mean}$ across those redshift ranges \citep[][]{Chang2018}. It is thus reasonable to compare our results with these studies despite the forward evolution of our simulation. 

\subsection{Residual trends}\label{sec:disc_residuals}
We find a significant residual trend towards smaller splashback radius as galaxy mass increases, even without dynamical friction (see Figure \ref{fig:rsp_distributions}). The source of this trend is unclear, but it impacts primarily the lowest-mass clusters. Because our fit quality for the bootstrapped profiles in this mass bin is in fact better than that of higher mass clusters (see the discussion in Appendix \ref{sec:appendixa}), we do not believe it results from a fitting issue. It is possible that the high-mass galaxies trace a different dynamical path than other galaxies, which appears more prominently among the lower-mass clusters. 

Our galaxy samples are also greatly affected by the output of the TNG clusters at $z=0$, as they form the initial conditions of our simulations. \citet{ONeil2022} worked with the same sample of haloes and subhaloes, and found similar, but larger, trends at $z=0$ (0~Gyr). That the magnitude of the residual trend we measure has decreased relative to \citet{ONeil2022} suggests that the trend may arise from a non-DF process we do not model (e.g. baryonic physics). 

We have examined the ``contamination'' of our sample at 5~Gyr (i.e. what fraction of galaxies within $2R_\mathrm{200,mean}$ at 5~Gyr were inside that radius at the start of our simulation), to see how these galaxies, which primarily trace the splashback radius, might be more or less affected by the starting conditions. However, this contamination fraction is relatively constant across cluster masses. Consequently, we cannot conclude that the $R_\mathrm{sp}$ measurements of lower mass clusters are differently impacted by the simulation starting conditions than high mass clusters. Instead, we conclude that massive galaxies trace a different path in the phase-space, particularly in low-mass clusters. We do not explore this difference further, though it is deserving of more study in the context of the splashback radius \citep[see e.g.][for more discussion on galaxy mass and phase-space]{Rhee2017, Sampaio2024}.

\subsection{Dynamical friction implementation}\label{sec:disc_df_implementations}
The use of these idealized simulations, and the analytical implementation of dynamical friction, allows us to uniquely isolate DF. However, the analytical implementation of DF includes approximations not present in an N-body approach. Appendix \ref{sec:appendixb} introduces a limited parameter study of the Coulomb logarithm $\Lambda$, an uncertain term in Chandrasekhar's analytical DF formalism. We find that the uncertainties in the Coulomb logarithm do not significantly affect the location of the splashback radius (measured by subhalo positions). High-mass clusters are even less responsive to adjustments in $\Lambda$.

We also note that Chandrasekhar's dynamical friction formula is thought to overestimate the role of DF compared to N-body simulations \citep[see the merging timescales found by][]{BoylanKolchin2008, Jiang2008}, again suggesting that our results are an upper limit on the strength of dynamical friction.

\subsection{Subhalo mass fraction requirement}\label{sec:disc_subhalo_stellar_mass}
In the literature, a variety of approaches, typically in simulation, have been used to find the minimum mass fraction ($M_{\rm subhalo}/M_{\rm halo}$) that is required for dynamical friction to have a significant impact on the subhalo orbit, which we expect will correspond to a change in the splashback radius. Though the subhalo mass is the dynamically relevant quantity, we express our results in terms of stellar mass because it is far more relevant for observations. 

In order to compare to the literature we must understand how the stellar mass and subhalo mass are connected. The stellar mass-halo mass relation has been broadly studied \citep[e.g.]{ Mitchell2016} and examined in IllustrisTNG as well \citep{Pillepich2018b}. With our sample of galaxies, we find that stellar mass cuts correspond to median subhalo masses as outlined in Table \ref{tab:stellar_halomass}, which are consistent with results from \citet{ONeil2024}. There is a slight dependence of the subhalo mass to stellar mass on cluster mass, with $M_\mathrm{subhalo}/M_*$ as a function of stellar mass decreasing for greater cluster mass. However, the trend is consistent between cluster masses ($M_\mathrm{subhalo} \propto M_*^{0.5}$), and are all well within 1$\sigma$, so only the overall median is presented here.

Work by \citet{vandenBosch1999} performed N-body simulations of an isothermal halo with Milky Way-like mass $\sim 10^{12} ~\mathrm{M_\odot}$, and studied eccentricities of orbiting objects. They found DF timescales to be Gyr-scale, and that DF is negligible for satellites with mass under $10^9~ \mathrm{M_\odot}$, producing a mass fraction requirement of $M_{\mathrm{subhalo}}/M_{\mathrm{halo}}~>~0.1\%$. They argue that this 0.1\% figure scales for galaxy clusters, while assuming that cluster timescales are equivalent to the Milky Way simulation timescale. Therefore they expect dynamical friction to matter only for subhaloes of total mass $> 10^{12} ~\mathrm{M_\odot}$, within a hypothetical cluster of mass $10^{15}~ \mathrm{M_\odot}$. A subhalo mass of $10^{12} ~\mathrm{M_\odot}$ corresponds to a stellar mass of $\sim2 \times 10^{10} ~\mathrm{M_\odot}$ (see Table \ref{tab:stellar_halomass}). Using a cut of $M_* > 10^{10} ~\mathrm{M_\odot}$ for our most massive cluster bin, we do not see a significant reduction in $R_\mathrm{sp}$, suggesting that a mass fraction above 0.1\% may be required.

More recent work by \citet{vandenBosch2016}, however, argues that a mass ratio of 10\% is necessary for dynamical friction to notably alter an orbit, among haloes of mass $\gtrsim10^{13} ~\mathrm{M_\odot}$. They examine three different cosmological simulations (from the Bolshoi and Chinchilla suites) with varying cosmological parameters and mass resolutions. They find this mass fraction requirement across all three simulations. The 10\% figure is consistent for our findings in the high cluster-mass regime. For example, among our most massive clusters, ($\sim 10^{15} ~\mathrm{M_\odot}$) we would expect DF to be impactful only for stellar mass cuts exceeding $7 \times 10^{11} ~\mathrm{M_\odot}$. We thus would not expect to find significant differences for any of the stellar mass cuts shown in Figure \ref{fig:rsp_distributions}, and we do not. On the other end, for a cluster of mass  $10^{13} ~\mathrm{M_\odot}$ in our sample, we would require a minmum stellar mass of $\sim 2 \times 10^{10} ~\mathrm{M_\odot}$ for DF to have an impact. We do see a notable ($\gtrsim 1\sigma$) reduction in the splashback radius for a cut of $M_* > 10^{10} ~\mathrm{M_\odot}$ after 5~Gyr, suggesting that the requirement of \citet{vandenBosch2016} may be too stringent. However, direct comparison is challenging because they did not divide their sample by halo mass. They, additionally, account for processes such as tidal stripping that contribute to subhalo mass loss, which are excluded from our study. Consistent with our comparison, it is expected that a higher starting mass fraction would be required if mass loss processes are included, because mass loss should reduce the strength of dynamical friction.

Yet other studies have sought a middle ground, finding that a mass fraction of 1\% is a suitable requirement for dynamical friction to have an impact on subhaloes \citep{Adhikari2016,More2016}. \citet{Adhikari2016} examined clusters of virial mass $\geq 10^{14}$, and subhaloes with radial plunging orbits. They found the splashback radius was reduced by up to $\sim 20\%$ for low-mass clusters ($3 - 5 \times 10^{13} ~\mathrm{M_\odot}$) between subhaloes with masses less than- versus exceeding $4\times10^{11} ~\mathrm{M_\odot}$, which corresponds to a stellar mass cut of $\sim 3 \times 10^9~ \mathrm{M_\odot}$. In the matching cluster mass bin ($10^{13.5} \leq M_\mathrm{200,mean}/\mathrm{M_\odot} < 10^{14}$), we find only a 7\% reduction in the splashback radius, even using a stellar mass cut of $10^{10}~\mathrm{M_\odot}$. Plunging radial orbits \citep[as used in][]{Adhikari2016} should introduce the most extreme case of DF, so it is unsurprising that their subhaloes are more affected than those of this simulation. They also find that DF has a reduced effect in lower-concentration clusters, bringing their results closer to ours. Future work could examine the role of cluster concentration with the IllustrisTNG sample, using our more physically motivated subhalo orbits.

The wide variation in literature mass fractions reinforces the need for numerical simulations to ascertain the complex dependence of DF on subhalo mass, size, position, velocity, as well as halo density. We generally find that requiring $M_{\mathrm{subhalo}}/M_{\mathrm{halo}}~>~1\%$ captures our results, though again, we argue that $M_{\mathrm{*}}/M_{\mathrm{halo}}$ can be a more useful metric. For clusters of mass greater than $10^{14} \mathrm{M_\odot}$, a reduced splashback radius should be traced only by galaxies with stellar masses exceeding $10^{10}$~M$_\odot$, consistent with Figure \ref{fig:rsp_distributions}. 

\subsection{Splashback radius measurements}\label{sec:disc_rspratio}
Compared to previous simulations, Fig. \ref{fig:rsp_ratio_plot} suggests that dynamical friction alone cannot account for the discrepancies found in \citet{McAlpine2022}, who use a stellar mass cut of $10^8~\mathrm{M}_\odot$ and see a $\sim10\%$ reduction in splashback radius, or \citet{ONeil2021}, who found a $\sim12\%$ discrepancy between dark matter and galaxy splashback radii for cluster masses of $10^{13} - 10^{13.5}~\mathrm{M}_\odot$, but used a subhalo \textit{total} mass cut of $10^9~\mathrm{M}_\odot$. \citet{Diemer2017} state that dynamical friction can reduce splashback radius measurements up to 20\%, which exceeds the maximal reduction that we find.

Magnitude cuts in surveys will preferentially select more luminous galaxies that are expected to be more massive, which may enhance the impact of dynamical friction. Still, most observing projects use cluster masses above $10^{14} ~ \rm M_\odot$ \citep[e.g.]{More2016, Chang2018, Shin2019}. For such surveys, regardless of the magnitude limitations, dynamical friction should not significantly influence the measured splashback radius. For example, \citet{Chang2018} uses an apparent magnitude cut of $i =21.5$, which at $z=0.55$ corresponds to an $i$-band absolute magnitude of -21. Using the magnitude-mass diagram for TNG galaxies, the \citet{Chang2018} sample has a minimum stellar mass of $10^{10} ~ \mathrm{M_\odot}$ \citep{ONeil2021}. Even with such high stellar masses, for these halo masses ($\left\langle M_\mathrm{200,mean} \right\rangle = 2.5 \times 10^{14} \mathrm{M_\odot}$), DF does not produce a >1$\sigma$ effect on the splashback radius.

Despite the large number of recent studies on the splashback radius, the question of why galaxies trace smaller $R_\mathrm{sp}$ values than the dark matter has remained unclear. Many studies have attributed this lag to DF, but few have investigated the process itself. Our simulations, idealized though they are, provide a unique window into the behavior of dynamical friction in the cluster environment. The uncertainties inherent to our methodology will enhance the impact of dynamical friction (e.g. the dynamical friction formula, the lack of tidal stripping), suggesting that dynamical friction is even weaker than suggested by our findings here. We thus find that dynamical friction is an insufficient answer to these discrepancies. Future studies of the splashback radius are challenged to develop a more comprehensive explanation for the splashback radius discrepancies \citep[whether it be due to selection effects or limitations in the splashback radius profile fitting, as outlined by][respectively]{ONeil2022, Diemer24}.

\section{Conclusions}\label{sec:conclusions}

Values of the splashback radius, when measured through galaxy number density, have been found to be smaller than splashback radius measurements made from dark matter profiles (on $\gtrsim 10 $\% scales). This trend persists between observations and simulations. Dynamical friction has been proposed as a force that would uniquely impact the orbits of galaxies, without affecting the dominant, but diffuse, background components of the halo. In contrast to past work, we argue that dynamical friction does not significantly reduce the splashback radius, aside from exceptional circumstances.

Here, we present the results of idealized simulations of galaxy clusters, with dynamical friction implemented analytically through Chandrasekhar's formula \citep{Chandrasekhar1943}, in order to examine the impact of dynamical friction. The cluster gravitational potentials and subhaloes are initially drawn from the IllustrisTNG cosmological simulation at $z=0$ to ensure that their properties (e.g. phase-space distribution of subhaloes, mass distribution, etc.) are relatively physical. Subhaloes were then evolved for 10~Gyr twice: both within the cluster's gravitational potential, and once with the analytical dynamical friction force. We thus isolate dynamical friction, and focus on how the presence of dynamical friction affects the splashback radius of the cluster when measured through the subhalo number density. In particular, we find that:
\begin{itemize}
    \item Dynamical friction shifts the location of the observed splashback radius, $R_{\mathrm{sp}}$, inwards. We find a \textit{maximal} reduction of the splashback radius because of DF to be approximately 10\%. However, we do not expect measurements of $R_{\mathrm{sp}}$ to be significantly shifted for clusters of $M_{\mathrm{200,mean}} \geq 10^{14} \mathrm{M_\odot}$. Therefore many  observational measurements of the splashback radius that find lower-than-expected values of $R_{\mathrm{sp}}$ should not attribute discrepancies to dynamical friction.
    \item The impact of DF on the splashback radius increases, as expected, with the fraction of subhalo (stellar) mass to halo mass (see Fig. \ref{fig:rsp_ratio_plot}), as opposed to the subhalo mass, reinforcing the findings of \citet{ONeil2022} that dynamical friction can not be the sole source of disagreements between dark matter and galaxy number density profiles.
    \item Dynamical friction has a greatest effect on galaxies with longer residence times. To what degree these galaxies contribute to the splashback feature partially determines the influence of DF. Among low-mass clusters (e.g. $10^{13} < M_\mathrm{200,mean}/\mathrm{M_\odot} < 10^{13.5}$), the splashback radius is best traced by galaxies with residences times exceeding 7~Gyr, further reducing the measured $R_\mathrm{sp}$.
    
\end{itemize}

We find that dynamical friction cannot alone account for the many reported cases of reduced splashback radii when measured with galaxy number density (i.e. rather than dark matter density), particularly among the highest-mass clusters that are most commonly studied \citep[see e.g.][]{More2016,Chang2018, Shin2019}.

The semi-idealized simulations presented here exclude other processes that affect the orbits of subhaloes inside the cluster and alter the masses of these subhaloes. For example, processes such as tidal stripping, if they were included, are expected to weaken the impact of dynamical friction as they reduce the masses of subhaloes during infall \citep[e.g.]{vandenBosch2016}. Our use of Chandrasekhar's analytical formula may also overestimate the strength of the dynamical friction \citep{BoylanKolchin2008, Jiang2008}. Each of these factors would further reduce the contribution of DF to splashback radius discrepancies measured in observation or from cosmological simulations.

This investigation targets isolated clusters at $z=0$. It is well-known that DF plays a significant role in halo mergers, which we do not study here. Measurements of halo splashback radii at higher redshift, with higher concentrations \citep[e.g.][]{Adhikari2016} or among less relaxed clusters, may be more strongly affected by dynamical friction. Future studies could investigate these parameters, as well as increase the complexity of modeling.

\section*{Acknowledgements}
We thank the anonymous reviewer for their insightful comments, which contributed greatly to the clarity of the paper. The authors also thank Jo Bovy for their help with \textsc{galpy}, and for their rapid responses to questions raised on the \textsc{galpy} GitHub page. The authors also would like to thank Frank van den Bosch for productive discussions during the early stages of the project. TMO gratefully acknowledges support through the Beckman Scholars Program by the Arnold and Mabel Beckman Foundation. TMO also thanks the efforts of her Wellesley College advisor James Battat for his support throughout the research process.
JB acknowledges support from NSF grant AST-2153201.
%%%%%%%%%%%%%%%%%%%%%%%%%%%%%%%%%%%%%%%%%%%%%%%%%%
\section*{Data Availability}
The data and code underlying this work are available at: \url{https://dx.doi.org/10.5281/zenodo.11357381}. All data used for this work is from the publicly available TNG300-1 simulation, and can be found at \url{https://tng-project.org} \citep{Nelson2019}.
 


\bibliographystyle{mnras}
\bibliography{example} % if your bibtex file is called example.bib

%\begin{thebibliography}{99}
%\end{thebibliography}

%%%%%%%%%%%%%%%%%%%%%%%%%%%%%%%%%%%%%%%%%%%%%%%%%%

%%%%%%%%%%%%%%%%% APPENDICES %%%%%%%%%%%%%%%%%%%%%

\appendix
\section{Fitting procedure} \label{sec:appendixa}
To enable analysis of stacked clusters, we perform bootstrapping and fitting of the galaxy number density profiles as described in Section \ref{sec:methods_fitting}, based on the \cite{Diemer2014} profile. In this appendix, we explore two elements of fitting (bin widths, as well as the fit parameters) to ensure the rigor of our fitting methods.

\begin{figure}
    \centering
    \includegraphics{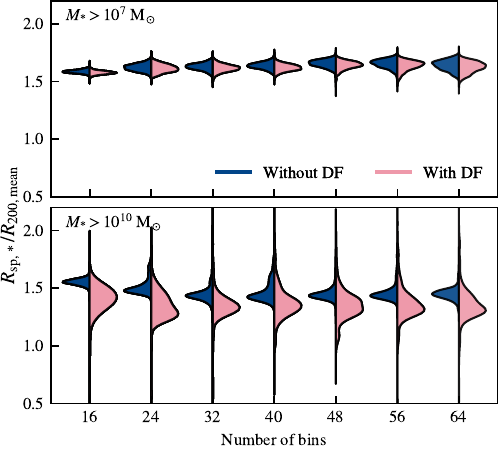}
    \caption{Violinplots of splashback radius distributions for a varying number of logarithmic radius bins, among haloes of $10^{13}~<~M_{200, \mathrm{mean}}~<~10^{13.5}~\mathrm{M}_{\odot}$, and for subhaloes of stellar mass greater than $10^7~\mathrm{M}_{\odot}$ (upper panel) and $10^{10}~\mathrm{M}_{\odot}$ (lower panel). The distributions evolved without DF are shown on the left in blue, while those with DF are shown on the right in pink.}
    \label{fig:violinplot}
\end{figure}

\subsection{Bin Widths}\label{sec:appendixa_binwidth}

When bootstrapping and fitting the density profiles (see Section \ref{sec:methods_fitting}), binning is an important criteria to consider. We use the standard logarithmically spaced bins to ensure that the inner bins are well-sampled. Too small a number of bins may wash out real variations in $R_{\rm sp}$ measurements, while the use of too many bins produces noisy data due to poorly sampled bins. Ideally, the splashback radius distributions should not vary with the number of bins used. Independence from binning indicates that the measurements are not residual effects of the binning choices, and instead reflect real variations in measurements of $R_{\rm sp}$. 

Various bin densities are used throughout the literature for splashback radius studies. Different sources cover a different radius range, but for studying $R_\mathrm{sp}$, they typically reach out to $\sim$ 5-10 $R_\mathrm{200,mean}$, ranging from 15 logarithmically spaced bins \citep{Murata2020,Dacunha2022}, to 32 \citep{ONeil2022}, 40 \citep{McAlpine2022}, and 42 (linear spacing) \citep{ONeil2021}.

In Figure \ref{fig:violinplot} we show a violinplot of the splashback radius distribution as a function of number of bins used in the bootstrapping process, to check the resilience of our results to binning. In the top panel, including all galaxies with stellar mass above $10^7~\mathrm{M}_\odot$, the distributions are narrow and similar between the different binning options (i.e. left to right), indicating we are accurately measuring the splashback radius. For the bottom panel, the sample of subhaloes is greatly reduced, producing greater error in the bootstrapping and fitting processes. The cluster subset includes 770 haloes, but of the $3.3~\times~10^6$ subhaloes with stellar masses above $10^7~\rm{M}_\odot$, only approximately 21,000 have $M_* > 10^{10}~\mathrm{M}_\odot$. We selected this subset of cluster and subhalo masses as they involved some of the most extreme density profile variation. The broader and more varied distributions reflect the greater uncertainties involved. Above 32 bins, the distributions appear relatively stable; however, in order to reduce noise in our data we use 32 bins for all data presented throughout.

\subsection{Fitting procedure}\label{sec:appendixa_fitting}
As described in Section \ref{sec:methods_fitting}, we fit the profile of \cite{Diemer2014} to our data, following, e.g., \cite{ONeil2022}. This fit involves nine parameters, as well as six demarcations between the inner, transition, and outer fit regions, which are not fit by the fitting implementation.

\begin{figure}
    \centering
    \includegraphics[width=\linewidth]{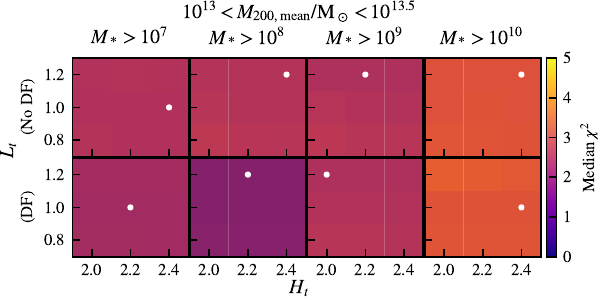}
    \includegraphics[width=\linewidth]{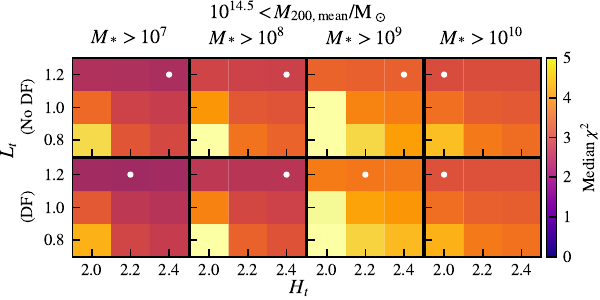}
    \caption{Median $\chi^2$ fit values for the lowest mass clusters (top) and highest mass clusters (bottom). For each cluster mass, fits are performed for each stellar mass cut (columns) and with and without dynamical friction (bottom and top rows respectively). This figure shows the parameter space of $L_t$ and $H_t$. White dots mark the parameters that minimize the median $\chi^2$ for each cluster and galaxy mass subsample.}
    \label{fig:chi_sq}
\end{figure}

\cite{Diemer24} identified issues with the \cite{Diemer2014} profile, including degeneracies among the parameters. However, we continue to use the \cite{Diemer2014} profile for consistency with earlier studies of the splashback radius. A further investigation of the fitting method is beyond the scope of this paper. Additionally, because of our idealized methodology, the splashback radius values evolve outwards to larger values than real splashback radii, which the profile was not designed for. Because we are interested in comparing the evolution of $R_\mathrm{sp}$ with and without DF, this limitation does not affect our analysis, but we must account for it during fitting. Thus, in this section, we present assessment of our fits.

We examine the parameter space of the inner (lower value $L_i$ and higher value $H_i$ such that the inner region is fit $L_i < r/R_\mathrm{200,mean} < H_i$), transition ($L_t$, $H_t$), and outer ($L_o$, $H_o$) fit regions, running fits on each selected combination for all cluster masses and performing chi-squared tests on each fit. We primarily focus our analysis on the transition region boundaries ($L_t$ and $H_t$) because those are the areas where we expect the splashback radius to have moved (compared to ``physical'' clusters for which the fit was derived).

Figure \ref{fig:chi_sq} shows the results of the chi-squared tests for our lowest and highest cluster samples, split by galaxy mass and the presence of DF to be consistent with the rest of our analysis. We present only the transition region parameters, $L_t$ and $H_t$ spaced by 0.2. The highest mass clusters ($>10^{14.5}$ M$_\odot$) are significantly more dependent on the fitting boundaries than other cluster masses (of which only one cluster mass is shown for comparison). For the high mass clusters, using $L_t = 1.2$ is produces significantly better fits. 

\begin{figure}
    \centering
    \includegraphics{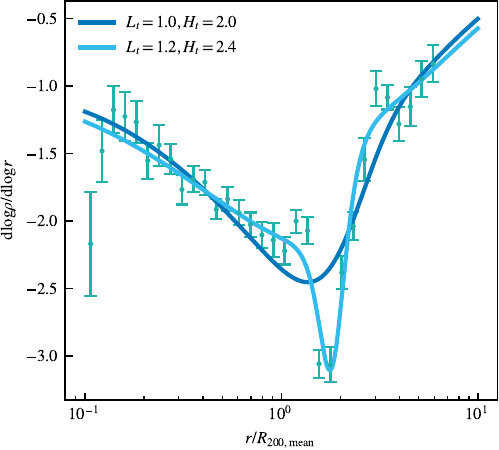}
    \caption{Fits (in light and dark blue) of data points (in green) for one example bootstrap from the highest mass ($>10^{14.5} \rm M_\odot$) cluster sample. The errors on the bootstrapped data are derived from sub-bootstrapping as per the methodology of Appendix B1 of \cite{ONeil2022}.}
    \label{fig:fit_example}
\end{figure}

Not every subsample indicates the same ``best-fitting" boundaries. Because the highest-mass clusters display the largest variation in $\chi^2$ due to the region boundaries, we choose to use $L_t = 1.2$ and $H_t = 2.4$ for consistency. Figure \ref{fig:fit_example} shows two example fits for a single bootstrap in the highest mass cluster sample, demonstrating how increasing $H_t$ improves the fit quality and captures a stronger dip in the differential density profile around the splashbak radius. We also use $H_i = 1.2$ \citep[as opposed to 1.0, as used by][]{ONeil2022} so that the inner and transition regions meet. Based on these tests, we thus fit all of the data presented in the main text with an inner region from $0.2 < r/R_\mathrm{200,mean} < 1.2$, a transition region from $1.2 < r/R_\mathrm{200,mean} < 2.4$, and an outer region $2.0 < r/R_\mathrm{200,mean} < 10.0$.

\section{Coulomb logarithm} \label{sec:appendixb}

As outlined in Section \ref{sec:methods_df_galpy}, the $\Lambda$ term in Chandrasekhar's analytical DF formula is not particularly well-defined. In the main body of text, we have used the default \textsc{galpy} implementation of $\Lambda$, which is computed as in Equation \eqref{eq:galpy_lambda}. Here, we investigate how strongly the $\Lambda$ term affects the dynamical friction term and the $R_{\mathrm{sp}}$ measurement. 

\citet{BoylanKolchin2008} and \citet{Jiang2008} found that merging timescales calculated from the analytical Chandrasekhar formulation are underestimated, relative to results from N-body and hydrodynamic numerical simulations, with \citet{Jiang2008} particularly citing the $\Lambda$ term as a source of uncertainty. Thus, we expect our results to, if anything, generally overestimate the impact of DF.

Some DF implementations instead utilize an unchanging $\Lambda$ related to the scale size of the halo, $R_{\rm halo}$. However such strategies less accurately match results from N-body simulations \citep{Just2005}; specifically, the use of a constant $\Lambda$ produces overly rapid inspirals and unphysical circularization of orbits \citep{Hashimoto2003}. Instead, \citet{Petts2015} use a maximum impact parameter of the subhalo's orbital radius, which forms the basis for the \textsc{galpy} implementation. \citet{Petts2015} note that this method can overly amplify the influence of dynamical friction.

Given these uncertainties, we examine the dependence of our results on $\Lambda$. We do not examine all possible variations of $\Lambda$ implementation, instead focusing on the use of a constant $\Lambda$ value. Past work has found that the use of a constant $\Lambda$ caused subhaloes to inspiral too quickly \citep{Hashimoto2003, Just2005}, which we might expect to reduce the splashback radius.

\begin{figure}
    \includegraphics{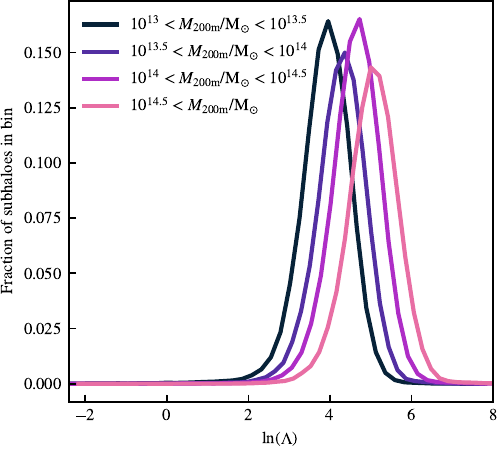}
    \caption{Distributions of $\mathrm{ln}(\Lambda)$ for all subhaloes within $R_{200, \mathrm{mean}}$ at the start of the simulation, grouped by halo mass. The histograms are scaled by the number of subhaloes.}
    \label{fig:lambda_distribution}
\end{figure}

The starting distributions of $\mathrm{ln}(\Lambda)$, as calculated in Equation \eqref{eq:galpy_lambda}, are shown in Figure \ref{fig:lambda_distribution}, for subhaloes beginning within $R_{200, \mathrm{mean}}$ at $z=0$. We select only these subhaloes because they are the ones most strongly influenced by dynamical friction (see e.g. Figure \ref{fig:radialvelocities}). We find that ln($\Lambda$) is normally distributed around ln$(\Lambda) \approx 4-5$. The trend towards greater $\Lambda$ among higher-mass clusters may be unintuitive; since we find DF to be less influential in those clusters, we might expect $\mathrm{ln}(\Lambda)$ to be smaller. However, the larger $\Lambda$ distributions among high mass cluster are reflective of the larger scale radii of the more massive clusters, i.e. $R_{200, \mathrm{mean}}$ is larger in more massive clusters. (Note again, we found that $r_{\rm hm} > GM/v^2$ for most galaxies in the simulation, so $r_{\rm hm}$ is the overwhelmingly the value used to calculate $\Lambda$, and it does not change greatly with halo mass.) The gravitational forces are simply far stronger in higher-mass clusters, reducing the relative importance of dynamical friction.

\begin{figure}
    \centering
    \includegraphics{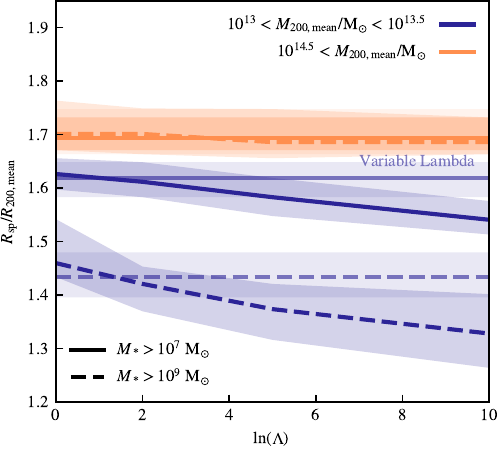}
    \caption{Median splashback radius distributions for constant $\Lambda$ systems, after 5~Gyr of evolution. Low and high mass haloes are included (purple and orange lines, respectively, as marked), with various stellar mass cuts (above $10^7$~M$_{\odot}$ in solid lines and $10^9$~M$_{\odot}$ dashed). Systems with no dynamical friction are shown at $\mathrm{ln}(\Lambda)$ = 0. The lightened lines show the position of the median splashback radius from the default variable implementation of dynamical friction (as used in Figures \ref{fig:profile_and_diff_dens_14.145} to \ref{fig:lambda_distribution}.) The shaded bands show the 16-84 percentile range for all data.}
    \label{fig:rsp_with_constant_lambda}
\end{figure}

To test the impact of $\Lambda$, we implement a set of constant-$\Lambda$ simulations. We evolve all haloes for 10~Gyr as before, but with fixed $\mathrm{ln}(\Lambda)$ values of 2, 5, and 10 for each subhalo. Figure \ref{fig:rsp_with_constant_lambda} shows how the splashback radius, measured again after 5~Gyr, evolves with $\Lambda$ for selected halo and subhalo masses. As expected, as $\mathrm{ln}(\Lambda)$ increases, the splashback radii decrease, especially for low-mass haloes, but barely exceeding 1$\sigma$. More massive haloes see little impact from adjustments of $\Lambda$, even when they exceed the median starting value by more than a factor of two. We attribute this to the relative weakness of DF compared to gravitational forces, such that significantly increasing DF has very little effect on the overall splashback distribution. Interestingly, even with a $\mathrm{ln}(\Lambda)=2$, which is smaller than the starting median value, the splashback radius is slightly reduced (though not significantly). That finding is consistent with previous work stating that using a constant $\Lambda$ contributes to overly rapid inspiral and enhanced dynamical friction \citep{Hashimoto2003, Just2005}. Nonetheless, our simple examination into the complexities and uncertainties in defining $\Lambda$, suggest that variations in $\Lambda$ cannot easily account for the splashback radius discrepancies seen in high-mass clusters.

\section{Subhalo angular momentum} \label{sec:appendixc}
Although a thorough examination of angular momentum in clusters is beyond the scope of our study, here we briefly discuss how dynamical friction affects the angular momentum of galaxies in our sample. 

\begin{figure}
    \centering
    \includegraphics{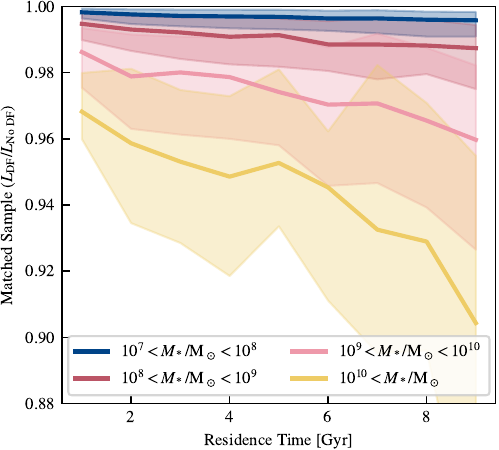}
    \caption{Median ratio of angular momentum of subhaloes after 10~Gyr evolved with and without dynamical friction, as a function of their residence time in the cluster. Here, we show only the least massive cluster bin, and separate galaxies by stellar mass. Note that we impose a maximum stellar mass cut as well as minimum, unlike the rest of the analysis.}
    \label{fig:angmom}
\end{figure}

Figure~\ref{fig:angmom} shows the ratio of angular momentum $L\equiv M_* r v_{\rm tan}$ for a matched sample of galaxies with and without dynamical friction. The sample consists of all galaxies that have the same residence time in both simulations. We expect to lose some galaxies due to variations in the infall time between the two simulations. In particular, this method is likely to exclude galaxies where the residence time is significantly altered by dynamical friction, potentially underestimating the impact of DF, however by comparing a matched sample, we are able to target dynamical friction directly.

The impact of dynamical friction on galaxy orbits may vary widely. For galaxies which pass very close to the cluster center, dynamical friction may radically affect their orbits and angular momentum. Figure~\ref{fig:prettypicture} shows a few extreme cases of these orbital changes. Most galaxies will experience a more moderated effect.  Fig.~\ref{fig:angmom} shows the population level impact of DF on angular momentum. 

As expected, more massive galaxies lose more angular momentum when dynamical friction is enabled. This decrease in $L$ and associated energy loss is also a function of residence time; galaxies that have resided in the cluster for longer have lost more angular momentum to dynamical friction. At its maximum, this effect reaches ${\sim}10\%$ for the most massive galaxies in the lowest mass clusters, consistent with our findings for the splashback radius in Fig.~\ref{fig:rsp_ratio_plot}. In the highest mass clusters ($M_{\rm 200,mean} >10^{14.5} \rm M_\odot$) the median galaxy does not lose more than 0.5\% of its angular momentum from dynamical friction. \\

%%%%%%%%%%%%%%%%%%%%%%%%%%%%%%%%%%%%%%%%%%%%%%%%%%

% Don't change these lines
%\bsp	% typesetting comment
\end{document}